\documentclass[twocolumn]{aastex62}
\usepackage{graphicx}
\usepackage{amssymb}
\usepackage{amsmath}
\usepackage{natbib}
\usepackage{todonotes} 
\usepackage{csvsimple}

\graphicspath{{./}{figures/}}

\shorttitle{Spectroscopic Extraction of Convective RVs}
\shortauthors{Miklos et al.}

\begin{document}

\title{Testing the Spectroscopic Extraction of Suppression of Convective Blueshift}


\author{M. Miklos}
\affiliation{Department of Physics, Harvard University, 17 Oxford Street, Cambridge MA 02138, USA}
\affiliation{Harvard-Smithsonian Center for Astrophysics, Cambridge, MA 02138, USA}
\author{T. W. Milbourne}
\affiliation{Department of Physics, Harvard University, 17 Oxford Street, Cambridge MA 02138, USA}
\affiliation{Harvard-Smithsonian Center for Astrophysics, Cambridge, MA 02138, USA}
\author{R. D. Haywood}
\altaffiliation{NASA Sagan Fellow}
\affiliation{Harvard-Smithsonian Center for Astrophysics, Cambridge, MA 02138, USA}
\author{D. F. Phillips}
\affiliation{Harvard-Smithsonian Center for Astrophysics, Cambridge, MA 02138, USA}
\author{S. H. Saar}
\affiliation{Harvard-Smithsonian Center for Astrophysics, Cambridge, MA 02138, USA}
\author{N. Meunier}
\affiliation{Univ. Grenoble Alpes, CNRS, IPAG, 38000 Grenoble, France}
\author{H. M. Cegla}
\altaffiliation{CHEOPS Fellow, SNSF NCCR-PlanetS}
\affiliation{Observatoire de Gen\`eve, Universit\'e de Gen\`eve, 51 chemin des Maillettes, 1290 Versoix, Switzerland}
\author{X. Dumusque}
\affiliation{Observatoire de Gen\`eve, Universit\'e de Gen\`eve, 51 chemin des Maillettes, 1290 Versoix, Switzerland}
\author{N. Langellier}
\affiliation{Department of Physics, Harvard University, 17 Oxford Street, Cambridge MA 02138, USA}
\affiliation{Harvard-Smithsonian Center for Astrophysics, Cambridge, MA 02138, USA}
\author{J. Maldonado}
\affiliation{INAF-Osservatorio Astronomico di Palermo, Piazza del Parlamento 1, 90134 Palermo, Italy}
\author{L. Malavolta}
\affiliation{INAF-Osservatorio Astronomico di Padova, Vicolo dell`Osservatorio 5, 35122 Padova, Italy}
\affiliation{Dipartimento di Fisica e Astronomia ``Galileo Galilei'', Universit{\`a} di Padova, Vicolo dell`Osservatorio 3, I-35122 Padova, Italy}
\author{A. Mortier}
\affiliation{Astrophysics Group, Cavendish Laboratory, University of Cambridge, J.J. Thomson Avenue, Cambridge CB 0HE, UK}

\author{S. Thompson}
\affiliation{Astrophysics Group, Cavendish Laboratory, University of Cambridge, J.J. Thomson Avenue, Cambridge CB 0HE, UK}
\author{C. A. Watson}
\affiliation{Astrophysics Research Centre, School of Mathematics and Physics, Queen's University Belfast, Belfast BT7 1NN, UK}
\author{M. Cecconi}
\affiliation{INAF-Fundacion Galileo Galilei, Rambla Jose Ana Fernandez Perez 7, E-38712 Brena Baja, Spain}
\author{R. Cosentino}
\affiliation{INAF-Fundacion Galileo Galilei, Rambla Jose Ana Fernandez Perez 7, E-38712 Brena Baja, Spain}
\author{A. Ghedina}
\affiliation{INAF-Fundacion Galileo Galilei, Rambla Jose Ana Fernandez Perez 7, E-38712 Brena Baja, Spain}
\author{C-H. Li}
\affiliation{Harvard-Smithsonian Center for Astrophysics, Cambridge, MA 02138, USA}
\author{M. L\'opez-Morales}
\affiliation{Harvard-Smithsonian Center for Astrophysics, Cambridge, MA 02138, USA}
\author{E. Molinari}
\affiliation{INAF-Fundacion Galileo Galilei, Rambla Jose Ana Fernandez Perez 7, E-38712 Brena Baja, Spain}
\affiliation{ INAF-Osservatorio Astronomico di Cagliari, Via della Scienza 5-09047 Selargius CA, Italy}
\author{Ennio Poretti}
\affiliation{INAF-Fundacion Galileo Galilei, Rambla Jose Ana Fernandez Perez 7, E-38712 Brena Baja, Spain}
\affiliation{INAF-Osservatorio Astronomico di Brera, Via E. Bianchi 46, 23807 Merate (LC), Italy}
\author{D. Sasselov}
\affiliation{Harvard-Smithsonian Center for Astrophysics, Cambridge, MA 02138, USA}
\author{A. Sozzetti}
\affiliation{INAF-Osservatorio Astrofisico di Torino, via Osservatorio 20, 10025 Pino Torinese, Italy}
\author{R. L. Walsworth}
\affiliation{Department of Physics, Harvard University, 17 Oxford Street, Cambridge MA 02138, USA}
\affiliation{Harvard-Smithsonian Center for Astrophysics, Cambridge, MA 02138, USA}

\begin{abstract}
Efforts to detect low-mass exoplanets using stellar radial velocities (RVs) are currently limited by magnetic photospheric activity. Suppression of convective blueshift is the dominant magnetic contribution to RV variability in low-activity Sun-like stars. Due to convective plasma motions, the magnitude of RV contributions from the suppression of convective blueshift is related to the depth of formation of photospheric spectral lines of a given species used to compute the RV time series. \cite{m2017Model, m2017Other}, used this relation to demonstrate a method for spectroscopic extraction of the suppression of convective blueshift in order to isolate RV contributions, including planetary RVs, that contribute equally to the timeseries for each spectral line. Here, we extract disk-integrated solar RVs from observations over a 2.5 year time span made with the solar telescope integrated with the HARPS-N spectrograph at the Telescopio Nazionale Galileo (La Palma, Canary Islands, Spain). We apply the methods outlined by \cite{m2017Model, m2017Other}. We are not, however, able to isolate physically meaningful contributions of the suppression of convective blueshift from this solar dataset, potentially because our dataset is from solar minimum when the suppression of convective blueshift may not sufficiently dominate activity contributions to RVs. This result indicates that,  for low-activity Sun-like stars, one must include additional RV contributions from activity sources  not considered in the Meunier et al. 2017 model at different timescales as well as instrumental variation in order to reach the sub-meter per second RV sensitivity necessary to detect low-mass planets in orbit around Sun-like stars.  

\end{abstract}





\section{Introduction}

The radial velocity (RV) method is the principal technique for constraining the masses of exoplanets \citep{FirstDetection_1995}. It provides complementary information to the transit method, e.g., as used by the \emph{Kepler} and TESS spacecraft \citep{Borucki977, 1538-3873-126-938-398, TESS2014} and ground-based transit surveys. The Keplerian reflex motion induced in a Sun-like star by an Earth-mass planet in the habitable zone is of order 10 cm~s$^{-1}$ \citep{Fischer2016}, the target sensitivity of next-generation spectrographs \citep{Pepe_et_al_2010}. However, contributions to observed stellar RVs from photospheric stellar activity often exceed 1 m~s$^{-1}$ even in the quietest Sun-like stars, posing a significant barrier to the detection of exoplanets by the RV method (e.g. \citealt{Saar_et_al_1997, schrijver_zwaan_2000, Isaacson_Fischer_2010, Motalebi_et_al_2015}). Several recent works describe a variety of models to mitigate the effects of magnetic activity on stellar RVs. One approach has been to study the Sun as a star, extracting solar activity estimates from images of the solar surface \citep{m2010, haywood, tim} and comparing to simultaneous disk-integrated spectral measurements. In order to reduce unwanted stellar signals from exoplanet searches, however, methods for extracting stellar activity directly from spectra, and not from ancillary datasets, must be developed.

For Sun-like stars with low activity, suppression of convective blueshift due to photospheric plage (hereafter $\rm RV_{\rm conv}$) dominates over the wavelength-independent photometric effects due to spots, or RV shifts induced by Earth-like exoplanets \citep{Meunier_et_al_2010_MDI, Dumusque_et_al_2014, haywood} (hereafter $\rm RV_{\rm sppl}$). \cite{m2017Model} (hereafter M17) have developed one model to isolate $\rm RV_{\rm conv}$ contributions based on the observed non-linear relationship between relative depths and absolute RV blueshifts of spectral lines of a given species (here neutral iron) driven by plasma flow in granules, as described in \cite{gray2009, reiners, m2017Other, GrayOostra2018}. The exact physical origin of this observed correlation is non-trivial: a correct description of spectral line formation necessitates the summation of many different line profiles, each formed at different depths in the photosphere, and requires a full three-dimensional treatment (e.g., see \citealt{Nordlund2009, Stein2012, Cegla_2013, Bergemann_et_al_2019} and references therein). An intuitive (though inexact) understanding of this relationship may be determined by considering a simplified 1D picture: in this model, rising plasma low in the photosphere exhibits strong RV blueshift, while plasma closer to the surface has most of its motion directed tangentially as it merges into intergranular lanes, thus exhibiting less RV blueshift \citep{Dravins_1981}. While many factors such as temperature, electron pressure, and atomic constants affect spectral line relative depth \citep{gray2005book}, for spectral lines of a given atomic species, line depth shows strong anti-correlation with height of formation in the stellar photosphere. Therefore, the absolute radial velocity blueshift shows a strong, non-linear relationship with line depth, commonly referred to as the third signature of stellar granulation \citep{gray2009}. M17 leverage the dominance of $\rm RV_{\rm conv}$ to write the RV time series derived from a set of lines $s_0$ as 
\begin{equation}
    \rm RV_0 = \rm RV_{\rm sppl} + \rm RV_{\rm conv}
\end{equation}
where RV${}_0$ is the radial velocity measured with this specific line list. $\rm RV_{\rm conv}$ are line-list dependent contributions due to the suppression of convective blueshift, and $\rm RV_{\rm sppl}$ are photometric variations (e.g. spots and plage), planetary signals, or other RV sources that are the same for all spectral lines. 

M17 makes use of the non-linear relationship between line depth and convective shift by writing an an RV time series from a sublist $s_1$ of $s_0$ with a restricted flux range can be written
\begin{equation}
    \rm RV_1 = \rm RV_{\rm sppl} +  \alpha \rm RV_{\rm conv}
\end{equation}
where $\alpha$ is the ratio of the weighted mean shift in radial velocity by suppression of convective blueshift from line list $s_1$ compared to line list $s_0$. Based on the third signature of granulation \citep{gray2009}, we would expect $\alpha < 1$ for a sublist $s_1$ comprising strong lines formed close to the top of the photosphere, and $\alpha > 1$ for a sublist of weak lines, formed deep in the photosphere. If a precise value for $\alpha$ is known or can be inferred, we can invert the observed $RV_0$ and $ RV_1$ time series to extract time series of interest $RV_{\rm conv}$ and $RV_{\rm sppl}$. 

Using time series $\rm RV_{\rm conv}$ and $\rm RV_{\rm sppl}$ extracted from solar photospheric images, M17 construct synthetic time series $\rm RV_0$ and $\rm RV_1$ using a value for $\alpha$ fitted from a solar atlas \citep{kurucz84, kurucz05} and added white noise. The authors then test several methods for estimating $\alpha$ on these synthetic time series, finding good convergence for the value of $\alpha$ across the methods (within 5\% of the true value for low-noise conditions) (M17). Using this calculated value of $\alpha$, they then recover and validate the original $\rm RV_{\rm sppl}$ time series. Ideally, this technique could be utilized to correct RV time series for $\rm RV_{\rm conv}$ contributions to lower the RV activity threshold. 

On real data, determining an absolute scale for radial velocities is challenging, making it difficult to precisely determine $\alpha$.  M17 apply these methods to HARPS exposures of HD207129 but find no agreement for values of $\alpha$ derived by different estimation methods, which they attribute to infrequent observations and low SNR. 

Using the solar telescope \citep{phillips16} operating with the HARPS-N spectrograph at the Telescopio Nazionale Galileo (TNG, \citealt{HARPSN_2012}), we extract high-resolution disk-integrated solar spectra \citep{dumusque}. We now have more than 50,000 high-SNR solar exposures spanning over 4 years of observing \citep{ACC}. In this work, we apply Meunier's methods to the first 2.5 years of the solar dataset (from Summer 2015 - November 2017) to attempt a recovery of a precise value of $\alpha$ for use in reconstructing $\rm RV_{\rm conv}$ and $\rm RV_{\rm sppl}$. In Section 2, we discuss our method for extracting line-by-line RVs from the HARPS-N solar spectra. In Section 3, we discuss various techniques for determining $\alpha$ from the resulting RV timeseries, and attempt to compute consistent $\alpha$ values. We conclude in Section 4 with a discussion of the resulting values, and possible explanations for why the model does not reduce RV RMS on our dataset

\section{Methods}
\subsection{Extracting RVs from HARPS-N spectra}
Third signature plots in the literature are often based on neutral iron lines to demonstrate the relationship between relative depth and absolute convective blueshift \citep{gray2009, reiners, GrayOostra2018}. In order to compare lines known to exhibit the third signature effect, we consider a line list from the NIST database for Fe I lines\footnote{\url{https://physics.nist.gov/PhysRefData/ASD/lines_form.html}}.  We extract disk-integrated HARPS-N solar spectra over a 2.5 year span, with an average of 51 exposures per day. We cut data taken in overcast weather, as identified using the HARPS-N exposure meter, and reject data for any day with five or fewer exposures. Each spectrum is shifted to a heliocentric reference frame using relative velocities from the JPL Horizons ephemeris  \citep{Horizons_1996}. We normalize the spectrum continuum by dividing by the corresponding blaze measurement - we propagate the photon shot noise error from each into the fits of each spectral line, as described below.

\begin{table}
\centering
\caption{\label{table:linelist} First ten iron line uses in this analysis. An extended line list containing all 765 spectral lines used in out analysis is available online at DOI:\dataset[10.5281/zenodo.3541149]{https://doi.org/10.5281/zenodo.3541149}.}
\begin{tabular}{l}
\hline \hline
Wavelength (\AA) \\
\hline
3922.91\\
3946.99\\
3948.10\\
3975.21\\
3995.98\\
4000.25\\
4000.46\\
4001.66\\
4022.74\\
4047.30\\
\vdots \\\hline
\end{tabular}
\end{table}

\begin{figure*}
\includegraphics[scale=.65]{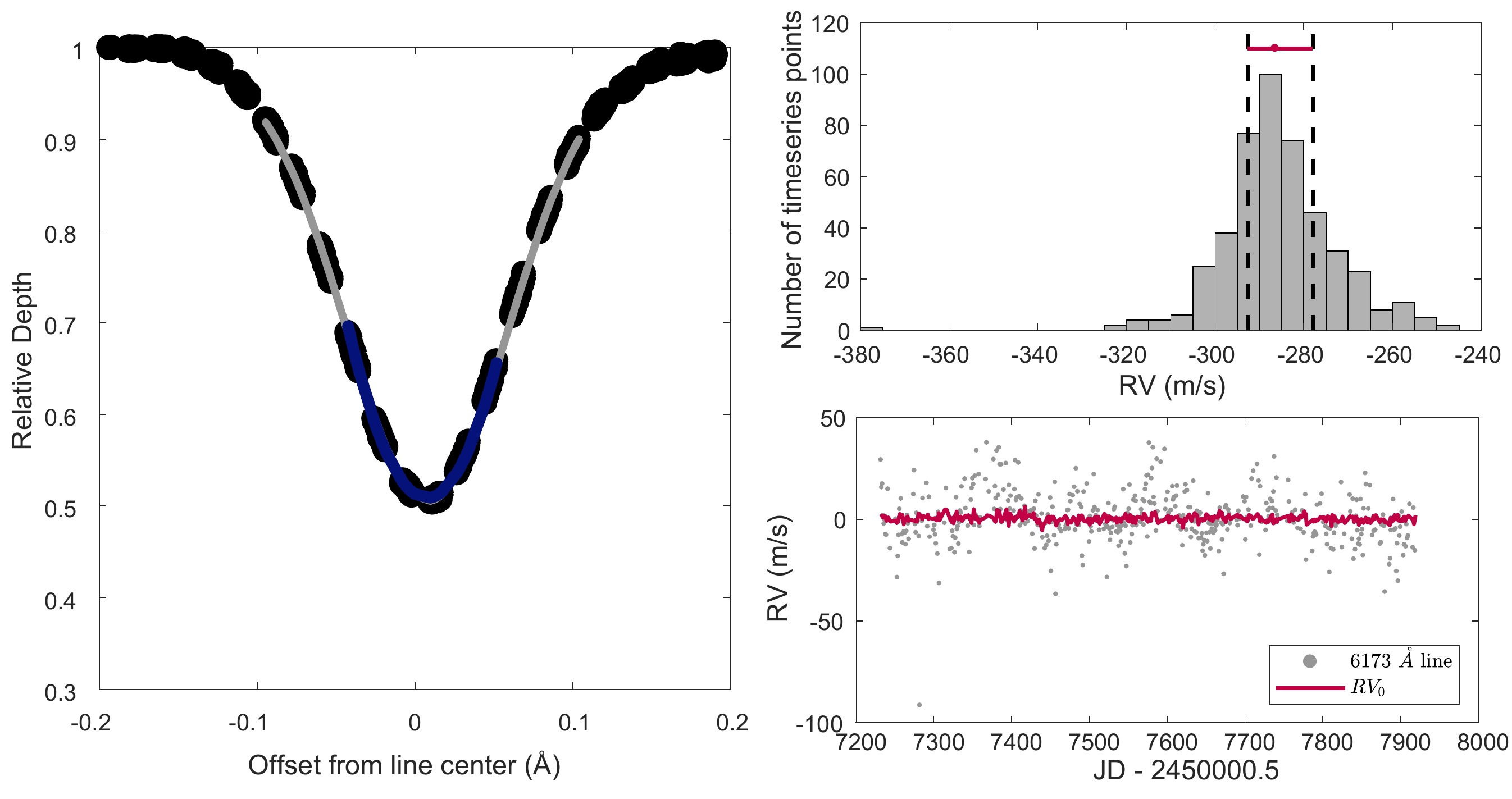}
\caption{Left: Illustration of boundaries of observed points (black) included in fits for Gaussian (gray curve) and polynomial (dark blue curve) fits for a representative line (at 6173 \AA). Right, top: demonstration of zeroing procedure for same 6173 \AA\,line-- the average of the middle two quartiles of RV values per line is subtracted off. Right, bottom: zeroed RV time series for 6173 \AA\,line (gray), compared to average time series for all lines, $RV_0$ (red).  \label{fig: fig1.png}}
\end{figure*}

For each individual spectral line, we stack the Doppler-shifted measurements from a given day to produce a composite line profile for each day. This approach, stacking data from different exposures instead of averaging over multiple exposures, avoids interpolating data onto a common wavelength grid. We fit the line core (0.2 Angstroms total) with Gaussian profiles to extract relative depth, and line center (0.1 Angstroms total) with 2nd-degree polynomials to measure the RV.\footnote{Full lists of each Gaussian fit parameter as a function of time for each spectral line are available online at DOI:\dataset[10.5281/zenodo.3541149]{https://doi.org/10.5281/zenodo.3541149}.} We adopted polynomial fits to best mirror the methods of M17. We then convert these line center positions from wavelength to RV. This process is illustrated in Figure \ref{fig: fig1.png}. Removing poorly-fit and blended lines results in a final list of 765 spectral lines, given in Table \ref{table:linelist}.

\begin{figure}
\centering
\includegraphics[scale=0.45]{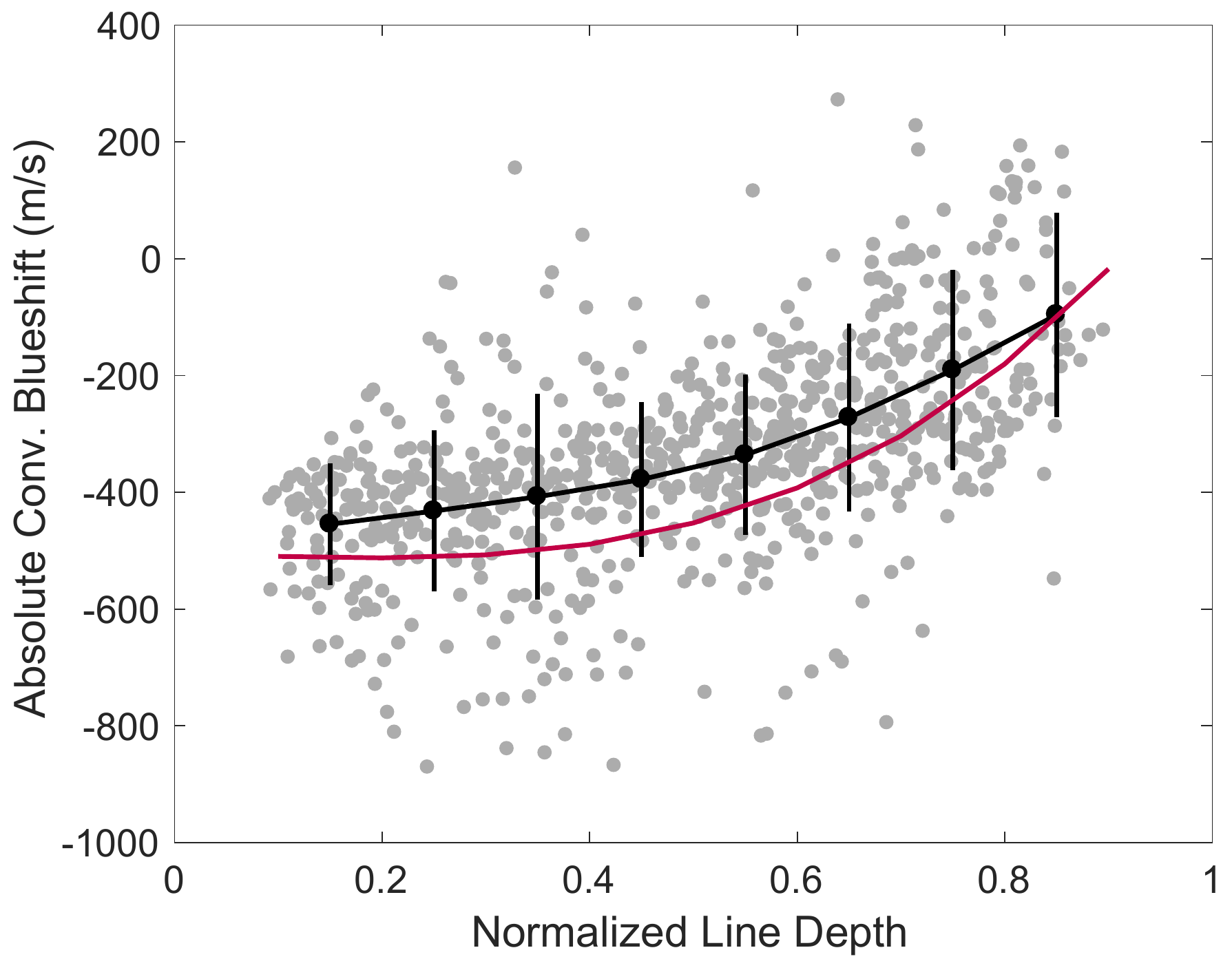}
\caption{Third signature of stellar granulation trend demonstrated in Fe I list extracted from NIST database. Lines are binned in 0.1 relative depth bins: black dots show the average value per bin, and errorbars show standard deviation per bin. The red curve shows the polynomial of best fit from \cite{reiners}.}
\label{fig: thirdSig}
\end{figure}

The relative velocities derived reproduce the shape of the third signature curve from \cite{reiners} up to an overall offset, as shown in Figure \ref{fig: thirdSig}. This offset may result from differences between the line lists or instruments used.

\subsection{Zeroing RV time series}
It is challenging to extract absolute RVs from spectral data. In accounting for blueshifts of individual spectral lines, we must identify the hypothetical RV value achieved in the absence of stellar activity, which will vary from line to line, in a manner that is robust against outlier points or noise. 
We zero the radial velocity time series per spectral line to account for this absolute blueshift. We sort time series observations by RV value per line, and subtract the  average of the middle two quartiles, as shown in Figure \ref{fig: fig1.png}. In selecting this range, we assume that low-activity days will fall close to the median value; by subtracting the average value for the low-activity days, we aim to identify the hypothetical no-activity point for each line while avoiding bias in our zero point due to outliers.

\subsection{Finding RVs from sublists}

Following the procedure of M17, we identify line sublists by relative depth. We take variance-weighted means of the entire line list ($s_0$), lines with relative depth .5-.95 ($s_1$), and lines with relative depth .05-.5 ($s_2$), to extract $\rm RV_0$, $\rm RV_1$, and $\rm RV_2$ respectively. The RV errors are computed from fit errorbars on the line center parameter, which incorporate propagated shot noise from the raw spectra. Features of these time series are listed in Table \ref{table:1}, while the time series themselves are given in Table \ref{table:RVlist}.

\begin{deluxetable}{lrrr}
\tabletypesize{\scriptsize}
\tablecaption{Features of RV time series extracted from HARPS-N/solar telescope daily binned spectra.\label{table:1}}
\tablewidth{0pt}
\tablehead{
\colhead{} & \colhead{$\rm RV_0$} & \colhead{$\rm RV_1$} & \colhead{$\rm RV_2$} 
}
\startdata
Relative Depth & .05-.95 & .5-.95 & .05-.5   \\
Number of Lines & 765 & 386 & 379\\
Standard Deviation (m~s$^{-1}$) & 1.50 & 1.64 & 1.74\\
Mean (m~s$^{-1}$) & .47 & .37 & .59\\
\enddata
\end{deluxetable}

\begin{table*}
\centering
\caption{\label{table:RVlist} The extracted time series $\rm RV_0$, $\rm RV_1$, and $\rm RV_2$ used in this analysis. The RVs derived from the HARPS-N DRS ($\rm RV_{DRS}$) are also provided as a point of comparison. The first ten RV values are given here - an extended list containing all values is available at DOI:\dataset[10.5281/zenodo.3541149]{https://doi.org/10.5281/zenodo.3541149}.}

\begin{tabular}{l l l l l}
\hline
\hline
JD - 2450000.5 & $\rm RV_0$ (m~s$^{-1}$) & $\rm RV_1$ (m~s$^{-1}$) &  $\rm RV_2$ (m~s$^{-1}$) & $\rm RV_{drs}$ (m~s$^{-1}$)\\
\hline
7232.51&1.34&1.26&1.46&5.66 \\
7233.54&2.00&2.15&1.81&6.71 \\
7234.51&0.96&-0.16&2.38&6.24 \\
7235.49&0.99&0.06&2.15&6.89 \\
7236.51&1.13&-0.34&2.99&7.48 \\
7237.49&1.79&0.68&3.19&7.12 \\
7238.56&0.59&-0.32&1.75&5.25 \\
7239.43&-0.70&-1.46&0.25&5.38 \\
7241.55&1.00&0.07&2.18&5.34 \\
7244.47&-1.08&-2.58&0.82&2.31 \\
\vdots & \vdots & \vdots & \vdots & \vdots \\
\hline
\end{tabular}
\end{table*}

\begin{figure}
\includegraphics[scale=.4]{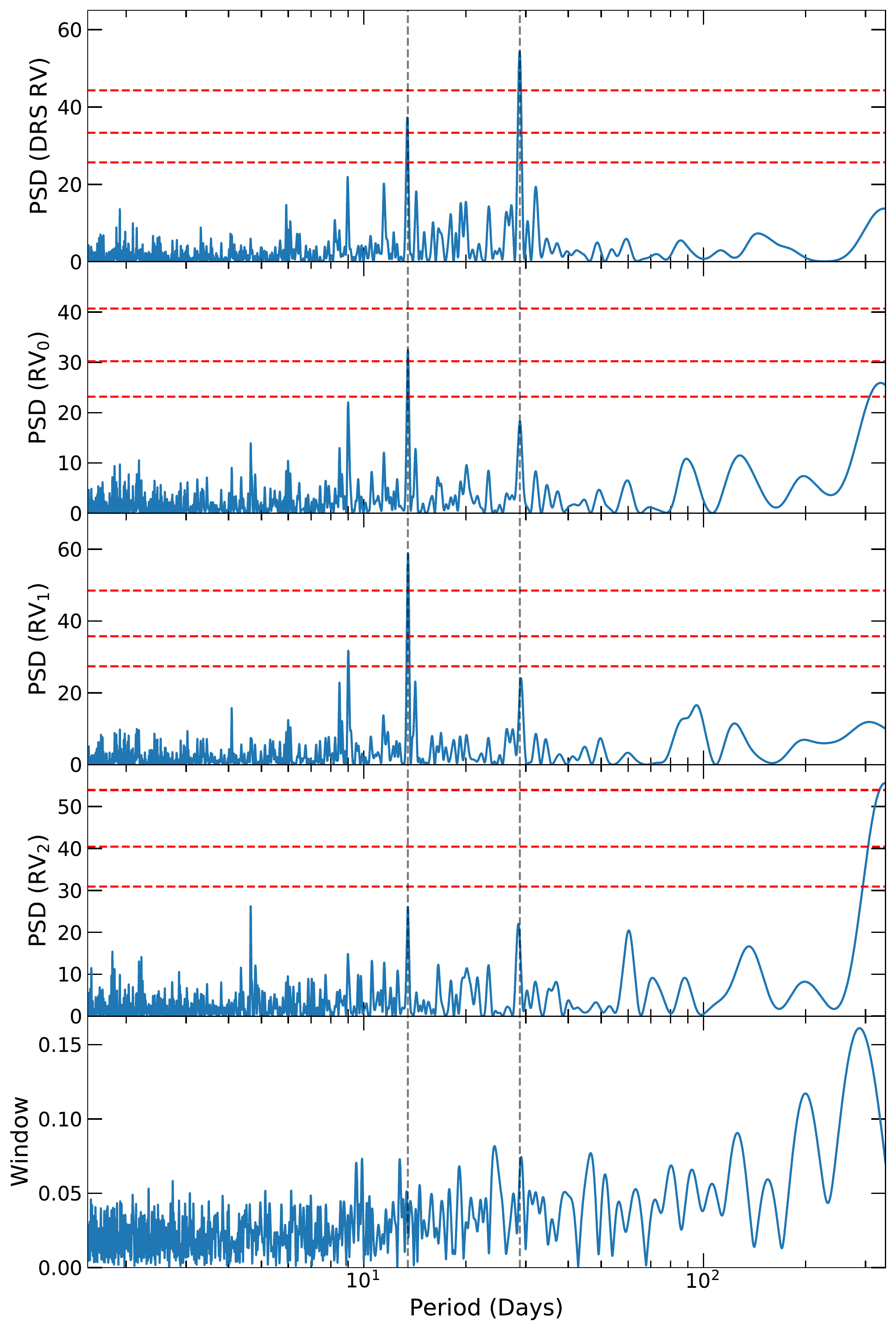}
\caption{Lomb-Scargle periodograms of the DRS-reduced RVs, as well as the $RV_0$, $RV_1$, and $RV_2$ time series generated from our line lists. The 10\%, 1\%, and 0.1\% False Alarm Probabilities (red dotted lines) are shown for each periodogram. The greatest power is concentrated in the solar synodic rotation period and its first harmonic (dotted gray lines), with no corresponding peaks in the window function (bottom panel).}
\label{fig: drsComp}
\end{figure}

To validate our extracted time series, we compare to the HARPS-N Data Reduction System (DRS) RVs \citep{Baranne_et_al_1996, Sosnowska_2012}. Figure \ref{fig: drsComp} shows a Lomb-Scargle periodogram comparison of the two time series \citep{GLS, VanderPlas2012, VanderPlas_2015}. Many of the periods with highest concentration of signal power align between the two series, suggesting that they capture the same solar physics. We note that the power concentrated at rotation and half-rotation periods in $\rm RV_1$ exceeds that of $\rm RV_2$, normalizing to the false alarm probability (FAP). While the RMS scatter of $\rm RV_2$ is greater than $\rm RV_1$, we find that over 1.5 $\rm m s^{-1}$ of white noise would need to be added to $\rm RV_2$ compared to $\rm RV_1$ to account for this difference in peak heights.  The model from M17 would predict, however, that since $\rm RV_2$ is calculated from lines presumably formed lower in the stellar photosphere, it should be more dominated by the suppression of convective blueshift, which would imply that this trend should be reversed. This observation presents the first suggestion that one of the assumptions that underlies the model is not realized on this dataset. 
\section{Analysis}
\subsection{Solving for $\alpha$}
If $\alpha$ is known, we can use experimentally determined values for $\rm RV_0$ and $\rm RV_1$ to extract theoretical time series of interest $\rm RV_{\rm conv}$ and $\rm RV_{\rm sppl}$. This process should isolate contributions from the suppression of convective blueshift, and leave behind common-mode planetary and photometric contributions in the corrected RV time series. In practice, however, we must estimate the parameter $\alpha$ by imposing assumptions on the reconstructed time series. We adopt methods to solve for the parameter $\alpha$ based on assumptions made for the reconstructed time series from M17. These methods rely predominantly on the assumption that $\rm RV_{\rm conv}$ dominates the RV time series. We applied the five methods detailed in M17 to solve numerically for the value of $\alpha$ that: 1) minimizes the mean absolute value of $\rm RV_{\rm sppl}$; 2) minimizes the correlation between $\rm RV_{\rm conv}$ and $\rm RV_{\rm sppl}$; 3) is the slope of $\rm RV_1$ vs $\rm RV_0$; 4) maximizes the ratio of the variance in $\rm RV_{\rm conv}$ vs that in $\rm RV_{\rm sppl}$; 5) maximizes that ratio when $\rm RV_{\rm sppl}$ is smoothed over 30 days, to average over rotationally modulated activity-induced variations. Additionally, 6) we calculate a best estimate for $\alpha$ as the ratio of the mean values of absolute RV time series derived from $\langle\rm RV_{\rm 1}\rangle/\langle\rm RV_{\rm 0}\rangle$ or $\langle\rm RV_{\rm 2}\rangle/\langle\rm RV_{\rm 0}\rangle$. 

\begin{deluxetable*}{lrrrr}
\tabletypesize{\scriptsize}
\tablecaption{Estimates for $\alpha$, and RMS variation for extracted $\rm RV_{\rm sppl}$ time series (m~s$^{-1}$) from different methods. Method 2 fails to converge likely due to correlated noise, as discussed in M17 \label{table: 2}}
\tablehead{
\colhead{Method} & \colhead{$\rm RV_1, \alpha$} & \colhead{$\rm RV_1, std(\rm RV_{\rm sppl})$} & \colhead{$\rm RV_2, \alpha$} &  \colhead{$\rm RV_2, std(\rm RV_{\rm sppl})$}
}
\startdata
1) $\langle\rm RV_{\rm sppl}\rangle = 0$ & .79 & 3.54 & 1.26 & 3.60 \\
2) Minimize correlation between $\rm RV_{\rm sppl}, \rm RV_{\rm conv}$ & N/A & N/A & N/A & N/A\\
3) Slope of $\rm RV_i$ vs $\rm RV_0$ & .97 & 22.63 & 1.03 & 28.54\\
4) Maximize $\rm std(\rm RV_{\rm conv})/\rm std(\rm RV_{\rm sppl})$ & .99 & 67.79 & 1.01 & 85.56\\
5) Maximize $\rm std(\rm RV_{\rm conv})/std(\rm RV_{\rm sppl}\rm \; smoothed \ 30 \ days )$ & .73 & 2.91 & 1.34 & 2.91 \\
6) $\langle\rm RV_{\rm i, abs}\rangle/\langle\rm RV_{\rm 0, abs}\rangle$& .74 & 2.99 & 1.32 & 3.05\\
\enddata
\end{deluxetable*}

\begin{figure}
\centering
\includegraphics[scale=.45]{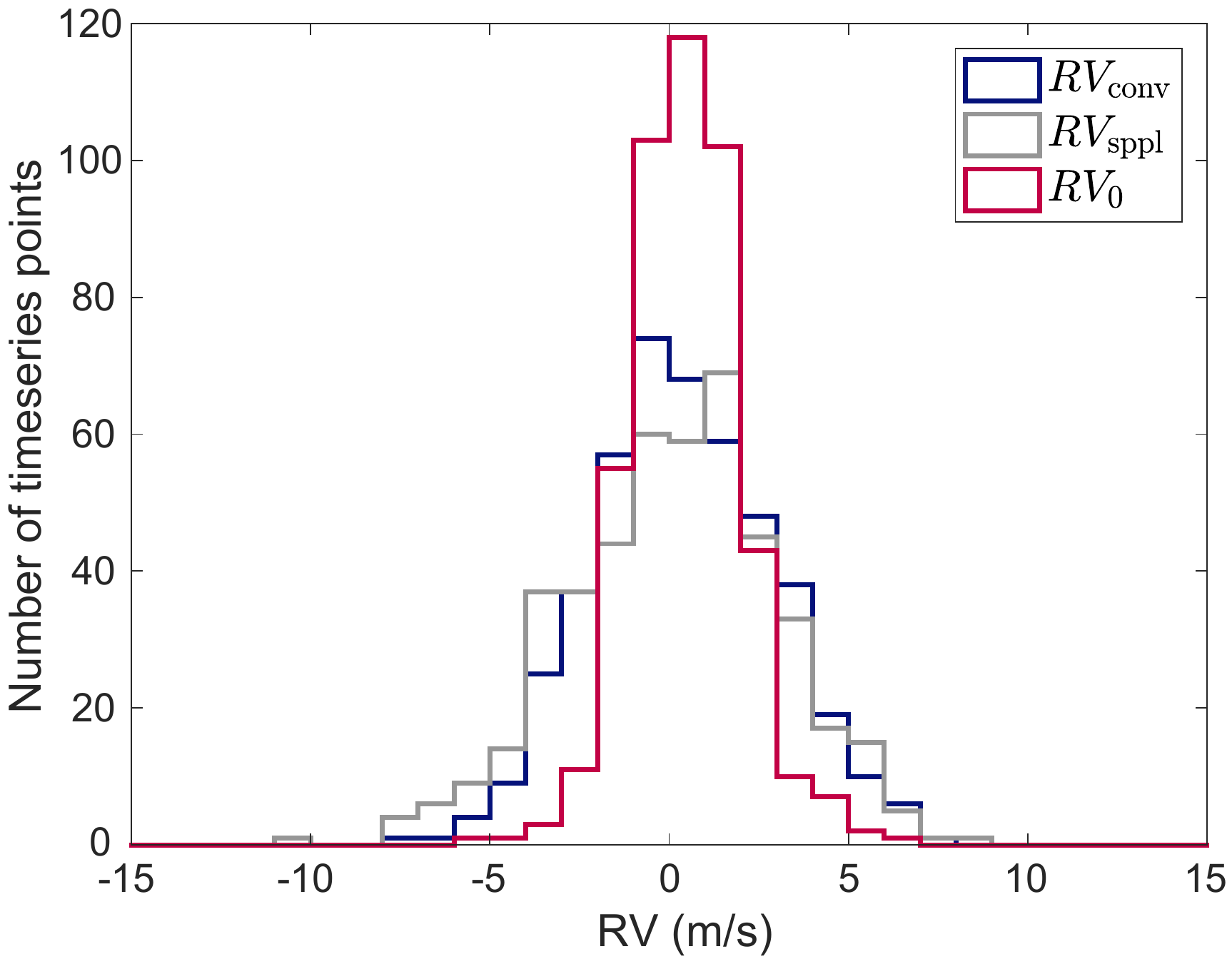}
\caption{Histogram of timeseries points of $\rm RV_0$, and treated timeseries $\rm RV_{\rm sppl}$, $\rm RV_{\rm sppl}$ reconstructed using $\rm RV_1$ for $\alpha = .73$: this reconstruction had the lowest RV RMS for $\rm RV_{\rm sppl}$. As shown, this reconstruction fails to reduce the RMS from the initial timeseries $\rm RV_0$. }
\label{fig:treated}
\end{figure}

These values are shown in Table \ref{table: 2}. Despite the relatively consistent observational sampling and high SNR of over 300 per exposure with an average of over 50 exposures per day, values for $\alpha$ differ based on the choice of assumption. Crucially, no physically motivated choice of $\alpha$ reduces the variability of $\rm RV_{\rm sppl}$, the corrected RV time series, compared to $\rm RV_0$, the untreated time series, as illustrated in Figure \ref{fig:treated}.

\section{Discussion}

Despite the higher SNR and better observational sampling for solar spectra, we are unable to extract physically significant reconstructed time series $\rm RV_{\rm conv}$ and $\rm RV_{\rm sppl}$ using the model and methods described in M17, suggesting that one of the assumptions of the model is not satisfied on this dataset. Crucially, the methods to estimate $\alpha$ assume that $\rm RV_{\rm conv}$ strongly dominates the RV timeseries. Potentially important is that our data set spans the activity minimum of the solar cycle, unlike the synthetic dataset from M17, which included a full solar activity cycle: the assumption that $\rm RV_{\rm conv}$ strongly dominates the RV time series might not hold for these restricted observations. Since the process of inverting the linear equations that describe $\left(\rm RV_0, \rm RV_1\right)$ to extract $\left(\rm RV_{\rm sppl}, \rm RV_{\rm conv}\right)$ amplifies all non-$\rm RV_{\rm conv}$ contributions, weakened $\rm RV_{\rm conv}$ at solar minimum could explain the inability to extract physical values of $\alpha$ on this data set. Our inability to reduce RV variability by applying the methods of M17 implies that sources of RV variability other than $\rm RV_{\rm conv}$ must be taken into account.

Additional results in the literature have shown that other processes besides $\rm RV_{\rm conv}$ may indeed play a dominant role near the solar minimum. For example, recent techniques demonstrate the ability to remove most power at the rotation period, but leave  1 m~s$^{-1}$ RV variability in corrected timeseries: \cite{Dumusque_2018} and \cite{Cretignier_2019} identify spectral lines insensitive to the suppression of convective blueshift; by computing RVs using these specially-selected line lists, the authors are able to reduce the RV RMS by a factor of 2.2, down to 0.9 m~s$^{-1}$. Independently, \cite{tim} use solar images from HMI/SDO to reproduce the activity-driven RVs. This analysis successfully removes the activity-driven signal at the rotation period, but still leaves an RMS amplitude of 1.2 m~s$^{-1}$. Using independent analysis frameworks, both techniques successfully remove the rotationally-modulated activity signal, but are still limited by some other processes. Other work is ongoing to characterize the contributions from granulation and supergranulation, which can contribute as much as 1 m~s$^{-1}$ to RV RMS \citep{Dumusque2011, Meunier2015, Meunier2019, Cegla2019}. The fact that our RV timeseries contain power concentrated at the rotation period and its harmonics is consistent with some significant $\rm RV_{\rm conv}$ contribution.\footnote{We note, however, that concentration of high-activity regions separated by 180 degrees longitude on the Sun containing not only plage but also long-lived sunspots that contribute to $\rm RV_{\rm sppl}$ through the photometric effect can also supply power at the rotation period \citep{schroeter, Shelke}; similar structure exists on other Sun-like stars \citep{Berdyugina2005}.} The inability to significantly reduce RV RMS using the methods of M17 makes sense in the context of  \cite{tim}, \cite{Dumusque_2018}, the literature on granulation and supergranulation, which demonstrate that well over 1 m/s of RV variation remains after accounting for $\rm RV_{\rm conv}$.

 When the linear equations defining $\rm RV_0$ and $\rm RV_1$ in terms of $\rm RV_{\rm sppl}$ and $\rm RV_{\rm conv}$ are inverted in the presence of noise introduced by this external variability, the RMS of noise in $\rm RV_{\rm sppl}$ is magnified to twice as large as the original RMS of noise in $\rm RV_0$ (M17). Potential contributions from instrumental systematics due to wavelength calibration \citep{HARPS-N_2014, Dumusque_2018, Cersullo_2019, Coffinet_2019}, or daily calibration sequences \citep{ACC}, may also contribute significantly to this non-$\rm RV_{\rm conv}$ RV variability. This sensitivity to RV contributions other than $\rm RV_{\rm conv}$ motivates future consideration of different solar activity processes, especially those operating on different timescales such as magnetoconvection \citep{Palle_et_al_1995, DelMoro_2004, Meunier_et_al_2018}. Furthermore, these additional processes, and even the suppression of convective blueshift itself, may contain subtle line list dependency, based on proxies for line responsiveness to magnetic activity such as the Lande-g factor (e.g., \citealt{Norton2006}). All of these contributions must be accounted for in order to reach the 10 cm~s$^{-1}$ detection limit of an Earth-like planet orbiting a Sun-like star. Future work is needed to identify correlates in spectra, solar images, or some other ancillary dataset that could be used to model these phenomenon. 

\acknowledgments
This work was supported in part by NASA award number NNX16AD42G and the Smithsonian Institution. Based on observations made with the Italian {\it Telescopio Nazionale Galileo} (TNG) operated by the {\it Fundaci\'on Galileo Galilei} (FGG) of the {\it Istituto Nazionale di Astrofisica} (INAF) at the  {\it Observatorio del Roque de los Muchachos} (La Palma, Canary Islands, Spain).  The solar telescope used in these observations was built and maintained with support from the Smithsonian Astrophysical Observatory, the Harvard Origins of Life Initiative, and the TNG.

This work was performed in part under contract with the California Institute of Technology (Caltech)/Jet Propulsion Laboratory (JPL) funded by NASA through the Sagan Fellowship Program executed by the NASA Exoplanet Science Institute (R.D.H.).

The authors thank A. Ravi for his assistance in preparing and submitting this manuscript.

A.C.C. acknowledges support from STFC consolidated grant number ST/M001296/1.

D.W.L. acknowledges partial support from the \emph{Kepler} mission under NASA Cooperative Agreement NNX13AB58A with the Smithsonian Astrophysical Observatory.

S.S. acknowledges support by NASA Heliophysics LWS grant NNX16AB79G.

H.M.C. acknowledges the financial support of the National Centre for Competence in Research PlanetS supported by the Swiss National Science Foundation (SNSF)

X. D. is grateful to the Branco-Weiss fellowship--Society in Science for continuous support.

This publication was made possible through the support of a grant from the John Templeton Foundation. The opinions expressed are those of the authors and do not necessarily reflect the views of the John Templeton Foundation.

This material is based upon work supported by the National Aeronautics and Space Administration under grants No. NNX15AC90G and NNX17AB59G issued through the Exoplanets Research Program. The research leading to these results has received funding from the European Union Seventh Framework Programme (FP7/2007-2013) under grant Agreement No. 313014 (ETAEARTH). 

This work was supported in part by the NSF-REU solar physics program at SAO, grant number AGS-1560313. 

The HARPS-N project has been funded by the Prodex Program of the Swiss Space Office (SSO), the Harvard University Origins of Life Initiative (HUOLI), the Scottish Universities Physics Alliance (SUPA), the University of Geneva, the Smithsonian Astrophysical Observatory (SAO), and the Italian National Astrophysical Institute (INAF), the University of St Andrews, Queen's University Belfast, and the University of Edinburgh.

This research has made use of NASA's Astrophysics Data System.

We thank the entire TNG staff for their continued support of the solar telescope project at HARPS-N.



\facility{TNG:HARPS-N}

\bibliographystyle{yahapj}

\end{document}